\documentclass[10pt]{article}
\usepackage{amssymb,amsmath} 
\usepackage{graphicx}
\usepackage[colorlinks=true, pdfstartview=FitV, linkcolor=blue, citecolor=blue, urlcolor=blue]{hyperref}
\usepackage{hyperref}

\def\div{\nabla\cdot } 
\def\pl{\partial}

\def\={\equiv}

\newcommand{\xt}{(\3x,t)}

\newcommand{\ox}{(\3x)}

\newtheorem{thm}{Theorem}

\newcommand{\bib}{\bibitem}

\newcommand{\nt}{\notag}
\newcommand{\ci}{\cite}
\newcommand{\lab}{\label}

\newcommand{\eq}{\eqref}

\newcommand{\hr}[1]{\href{#1}{\tt{#1}}}

\newcommand{\lp}{\left(}
\newcommand{\rp}{ \right)}
\newcommand{\lb}{ \left[}
\newcommand{\rb}{\right]}
\newcommand{\la}{\langle\,}

\newcommand{\ra}{\,\rangle}

\newcommand{\LB}{\left\lbrace}
\newcommand{\RB}{\right\rbrace}

\newcommand{\harr}[1]{\smash{\mathop{\hbox to .5in{\ \rightarrowfill\ }}
      \limits^{#1}}}

\newcommand{\0}[1]{{(#1)}}

\newcommand{\3}[1]{{\boldsymbol #1}}

\newcommand{\bh}[1]{{\boldsymbol{\hat #1}}}

\newcommand{\bbar}[1]{{\,\mathbb{\bar{ #1}}}}

\newcommand{\6}[1]{_{\scriptscriptstyle#1}}

\newcommand{\8}{\infty}
\newcommand{\9}[1]{^{\scriptscriptstyle#1}}

\def\c{\chi}
\def\d{\delta} 
\def\e{\varepsilon}

\def\l{\lambda} 
\def\m{\mu}

\def\p{\pi}

\def\r{\rho}
 
\def\s{{\sigma}}

\def\D{\Delta} 
\def\F{\Phi}

\newcommand{\db}{{\,{\rm d}\kern-1.6ex-}}
\newcommand{\dir}{{\pl\kern-1.2ex {/}}}
\newcommand{\dd}{{\rm d}}

\newcommand{\cc}[1]{{{\mathbb C\hskip.5pt}^{#1}}}

\newcommand{\curl}{\nabla\times}

\newcommand{\grad}{\nabla}

\newcommand{\ie}{{\it i.e., }}

\def\iff{\ \Leftrightarrow\ }
\newcommand{\im}{{\,\rm Im}\ }  
\newcommand{\imp}{\ \Rightarrow\ }
\newcommand{\inv}{^{-1}}

\newcommand{\lra}{\leftrightarrow}

\newcommand{\plra}{\pl^{\kern-1.25ex^\lra}}
\newcommand{\qq}{\quad} 
\newcommand{\qqq}{\qquad} 
\newcommand{\re}{{\,\rm Re}\  }   

\newcommand{\rr}[1]{{{\mathbb R}^{#1}}}

\newcommand{\supp}{{\rm supp \,}}

\def\XXint#1#2#3{{\setbox0=\hbox{$#1{#2#3}{\int}$}
     \vcenter{\hbox{$#2#3$}}\kern-.5\wd0}}

\def\bib#1{\bibitem[#1]{#1}}

\begin{document}

\title{Huygens' principle in classical electrodynamics:\\ a distributional approach}

\author{Gerald Kaiser\\
Signals \& Waves,  Austin, TX\\
\hr{http://www.wavelets.com}
}

\maketitle

\begin{abstract}\noindent
We derive Huygens' principle for electrodynamics in terms of 4-vector potentials defined as distributions supported on a surface surrounding the charge-current density. By combining the Pauli algebra with distribution theory, a compact and conceptually simple derivation of the Stratton-Chu and Kottler-Franz equations is obtained. These are extended to freely moving integration surfaces, so that the fields due to charge distributions in arbitrary motion are represented. A further generalization is obtained to multiple surfaces, which can be used to enclose clusters of transmitters, scatterers and receivers.
\end{abstract}

\tableofcontents

\section{Huygens' principle and communication}\label{S:PB reps}

The significance of Huygens' principle in physics has been described by Courant and Hilbert \ci[pp 765--766]{CH62} as follows:
\begin{quote}\noindent
\it ... our actual [3D] physical world, in which acoustic and electromagnetic (EM) signals are the basis for communication, seems to be singled out among other mathematically conceivable models by intrinsic simplicity and harmony. \rm
\end{quote}

Here is what they meant. Let $P\xt$ be the  \sl propagator \rm for the wave equation in $n$ space dimensions, which is the retarded solution of
\begin{align*}
\Box f\xt\=(c^{-2}\pl_t^2-\D_n)P\xt=\d\xt\=\d\0t\d\ox,\qq \3x\in\rr n,
\end{align*}
where $c$ is the propagation speed and $\D_n$ is the Laplacian in $\rr n$.
If a point source $\d\ox$ fixed at the origin is excited by a time signal $g\0t$, then the signal received at $\3x$ is the retarded solution of $\Box f\xt=g\0t\d\ox$, which is
\begin{align*}
f\xt=\int \dd\3x'\,\dd t'\, P(\3x-\3x', t-t')g(t')\d(\3x')=\int\dd t'\,g(t')P(t-t',\3x).
\end{align*}
The \sl ideal \rm communication is obtained only for $n=3$ since then
\begin{align*}
P\xt=\frac{\d(t-r/c)}{4\p r}\imp f\xt=\frac{g(t-r/c)}{4\p r},\qq r=|\3x|
\end{align*}
and $g\0t$ can be recovered directly from the received wave $f$. For $n=1$ and all \sl even \rm $n$, an impulse $g\0t=\d\0t$ produces a wave whose value at $\3x$ \sl reverberates \rm at times $t>r/c$. This can be seen in water waves ($n=2$), where the leading ripple is always followed by a train of ripples. In such a world, communication would require massive processing and information would generally be lost. For \sl odd \rm $n>3$, $P$ depends on $t$ through a sum of $\d(t-r/c)$ and its derivatives. This leads to a \sl distortion \rm of $g(t-r/c)$ by its derivatives. Thus ideal communication is possible only in a world with three spatial dimensions, as noted by Courant and Hilbert.

Huygens' principle in $\rr3$ is based on the above property of the wave equation. It states that the wave emitted by a source can be represented in the exterior of a surface $S$ surrounding the source region as a sum of secondary waves, called \sl Huygens wavelets, \rm emitted by points $\3x\in S$. Green's second theorem states that the Huygens wavelets consist of the propagators $P$ and their normal derivatives on $S$. In electrodynamics, Huygens' principle has been formulated in terms of the \sl Stratton-Chu equations \rm and the \sl Kottler-Franz equations. \rm The former give the exterior field in terms of the fields on $S$, while the latter give it in terms of the \sl tangential \rm fields on $S$. These equations have become a standard tool for analyzing EM scattering problems.

A generalization of Huygens' principle for scalar waves was derived in \ci{HK9} by letting $S$ be a sphere of radius $R$ and analytically continuing in $R$. This resulted in the deformation of the  Huygens wavelets emitted by the \sl points $\3x\in S$ to \sl pulsed-beam wavelets \rm emitted by \sl disks \rm $\5D$ tangent to $S$. The representation of radiation and scattering fields as superpositions of such pulsed beams has some attractive practical features. For example, the beams missing a given observer can be ignored without incurring a large error, and this gives an efficient method for numerical computation. A similar generalization was obtained in \ci{HK11} of Huygens' principle in electrodynamics, where the exterior field is represented as a sum of EM pulsed-beam wavelets.\footnote{Scalar (acoustic) and electromagnetic wavelets were introduced in \ci{K11} by analytically continuing solutions of the wave equation and Maxwell's equations to complex spacetime.
}
An added degree of numerical efficiency was gained by surrounding both the emitting and receiving sources by spheres and then analytically continuing in both radii.

I believe it is useful to study Huygens' principle in electrodynamics from fresh points of view in order to see how the above developments can be best applied and extended. In Section \ref{S:Pauli} we derive the \sl Pauli algebra \rm as the associative completion of the vector algebra in $\rr3$. This allows a well-known compact formulation of electrodynamics, as reviewed in Section \ref{S:EM}. In Section \ref{S:Huygens} we develop a generalized EM Huygens principle by synthesizing a global field from arbitrarily specified interior and exterior fields of a closed surface $S$. Distribution theory gives the required sources on $S$ radiating the given fields. In Section \ref{S:moving}, this is extended to arbitrarily moving surfaces. In Section \ref{S:part} we further generalize this scheme by synthesizing an EM field from its values in an arbitrary number of \sl cells \rm $\2E_k$ partitioning spacetime, with the appropriate sources on the interfaces between cells. This includes the two-sphere scheme in \ci{HK11}, whose cells consist of the interiors of the emission and reception spheres and the region between the two spheres.

\section{Derivation of the Pauli algebra}\lab{S:Pauli}

The equations of electrodynamics will be greatly simplified by using an associative algebra called \sl Pauli algebra \rm which can be regarded as a simple extension of the usual (non-associative) vector algebra in $\rr3$. Two vectors $\3A,\3B\in\rr3$ define a scalar product $\3A\cdot\3B$ and a vector product $\3A\times\3B$. We look for a product $\3A\3B$ consisting of a linear combination of these two bilinear expressions:
\begin{align}\lab{AB0}
\3A\3B=\3A\cdot\3B+\l\3A\times\3B,
\end{align}
where $\l$ is to be chosen so that the new product is \sl associative: \rm 
\begin{align}\lab{assoc}
(\3A\3B)\3C=\3A(\3B\3C)\=\3A\3B\3C.
\end{align}
Applying \eq{AB0} twice, we have
\begin{align*}
(\3A\3B)\3C&=\l(\3A\times\3B)\cdot\3C+(\3A\cdot\3B)\3C+\l^2(\3A\cdot\3C)\3B-\l^2(\3B\cdot\3C)\3A\\
\3A(\3B\3C)&=\l(\3A\times\3B)\cdot\3C+(\3B\cdot\3C)\3A+\l^2(\3A\cdot\3C)\3B-\l^2(\3A\cdot\3B)\3C.
\end{align*}
Thus \eq{assoc} is satisfied if and only if $\l^2=-1$. We arbitrarily choose $\l=i$ and define the \sl Pauli product \rm of $\3A$ and $\3B$ as
\begin{align}\lab{Pauli}
\3A\3B=\3A\cdot\3B+i\3A\times\3B.
\end{align}
The other choice $\l=-i$ is obtained by complex conjugation. For real vectors $\3A,\3B,\3C$, $\3A\3B$ is the sum of a real scalar and an imaginary axial vector
\begin{align}\lab{sv}
\la\3A\3B\ra_s=\3A\cdot\3B,\qq \la\3A\3B\ra_{\rm v}=i\3A\times\3B,
\end{align}
while $(\3A\3B)\3C$ is the sum of an imaginary scalar and a real vector. It follows that a \sl general \rm element of the Pauli algebra is the sum of a complex scalar and a complex vector, which we denote by
\begin{align}\lab{Agen}
\4A=A_0+\3A\ \ \hbox{with}\ \ A_0\in\4C,\ \3A\in\cc3.
\end{align}
The product of two such elements is then
\begin{align}\lab{4A}
\4A\4B=(A_0B_0+\3A\cdot\3B)+(A_0\3B+B_0\3A+i\3A\times\3B),
\end{align}
and \eq{assoc} implies that this, too, is associative:
\begin{align}\lab{assoc1}
(\!\4A\4B)\4C=\4A(\!\4B\4C)\=\4A\4B\4C.
\end{align}
A concrete representation of the algebra is given in terms of $2\times 2$ 
matrices by the correspondence \footnote{Along with complex numbers and quaternions, the Pauli algebra is one of the simplest examples of \sl Clifford algebra, \rm also known as \sl geometric algebra \rm \ci{H66,DL3}. Although its first application to physics was in quantum mechanics, it has also turned out to be useful in other fields, in particular classical electrodynamics \ci{B99}. The Pauli matrices $\s_k$ corresponds to the vectors $\bh x_k=\grad x_k.\ k=1,2,3$. 
}
\begin{align}\lab{rep}
\4A\lra\lb\begin{matrix}
A_0+A_3&A_1+iA_2\\ A_1\!-iA_2& A_0-A_3
\end{matrix}\rb,\qq A_0\in\4C,\ \3A=(A_1,A_2,A_3)\in\cc3,
\end{align}
with $\4A\4B$ represented by the matrix product. Under this correspondence, the Pauli algebra is therefore isomorphic to the algebra $GL(2,\4C)$ of all complex $2\times 2$ matrices.

\section{Application to electrodynamics}\lab{S:EM}

Consider electrodynamics in vacuum with Heaviside-Lorentz units $(\e_0=\m_0=1)$ and $c=1$. Define the spacetime differential operators
\begin{align}\lab{DDbar}
\4D=\pl_t-\grad \ \ \hbox{and}\ \  \bbar D=\pl_t+\grad,
\end{align}
which act on a Pauli-valued field $\4A\0x=\4A\xt$ on spacetime $\rr4$ by
\begin{align}\lab{del}
\4D\4A&=(\pl_t-\grad)(A_0+\3A)=(\pl_t A_0-\div\3A)+(\pl_t\3A-\grad A_0-i\curl\3A)\\
\bbar D\4A&=(\pl_t+\grad)(A_0+\3A)=(\pl_t A_0+\div\3A)+(\pl_t\3A+\grad A_0+i\curl\3A).\nt
\end{align}
Then
\begin{align*}
\4D\bbar D=\bbar D\4D=\pl_t^2-\grad^2\=\Box
\end{align*}
is the scalar wave operator.
Now consider the scalar wave equation
\begin{align}\lab{wave0}
\Box f\0x=g\0x
\end{align}
where $g\0x$ is a given source function which, for convenience, is assumed to be a distribution of compact support. The wave \sl radiated \rm by $g$ is the unique \sl causal\,\rm\footnote{In this context, causality simply means that $f$ is supported in the future region of $g$. If $g$ vanishes at $t=-\8$ as assumed here, it suffices to take the `initial condition' $f(\3x,-\8)=0$.
}
solution
\begin{align}\lab{sol0}
f\0x=\int_\rr4\dd^4x'\,P(x-x')g(x')=P*g\0x,
\end{align}
where $*$ denotes spacetime convolution and $P$ is the retarded propagator, which is the wave radiated by $g\0x=\d\0x\=\d\ox\d\0t$:
\begin{align}\lab{P}
P\0x=P\xt=\frac{\d(t-|\3x|)}{4\p|\3x|},\qq \Box P\0x=\d\0x.
\end{align}
Hence the wave operator is invertible on the space of such fields, with
\begin{align*}
\Box\inv=P*.
\end{align*}
Since $\Box$ is a scalar operator, it operates on Pauli fields $\4A\0x$ by
\begin{align*}
\Box\4A\0x=\Box A_0\0x+\Box\3A\0x.
\end{align*}
Thus we may extend the wave equation \eq{wave0} to Pauli fields as
\begin{align}\lab{wave1}
\Box\4F\0x=\4G\0x
\end{align}
where $\4G\0x$ is a Pauli-valued distribution with compact support. The unique causal solution is
\begin{align}\lab{sol1}
\4F\0x=P*\4G\0x=\int_\rr4\dd^4x'\,P(x-x')\4G(x').
\end{align}

We now apply the Pauli algebra to classical electrodynamics, more or less following \ci{B99}.  An EM field in free space consists of two vector fields $\3E\0x,\3B\0x$ satisfying Maxwell's equations
\begin{align}
&\pl_t\3E-\curl\3B=-\3J&&\div\3E=\r\lab{inhom}\\
&\pl_t\3B+\curl\3E=\30&&\div\3B=0\lab{hom}
\end{align}
where $(\r,\3J)$ is a given charge-current density. The obvious symmetry of these equations suggest combining the two fields into a single complex field
\begin{align}\lab{FEH}
\3F\0x=\3E\0x+i\3B\0x,
\end{align}
for which Maxwell's equations reduce to
\begin{align}\lab{Max}
&\pl_t\3F+i\curl\3F=-\3J && \div\3F=\r.
\end{align}
Now interpret $\3F\0x$ as a Pauli field with vanishing scalar component. Then \eq{del} shows that
\eq{Max} further reduces to the single equation
\begin{align}\lab{Max1}
\bbar D\3F=\r-\3J\=\4J.
\end{align}
The homogeneous equations \eq{hom} state that the source $\4J$ is real, but it will be useful to allow $\4J$ to be complex:
\begin{align}\lab{Jem}
\4J=\4J_e+i\4J_m&&\4J_e=\r_e-\3J_e&&\4J_m=\r_m-\3J_m
\end{align}
where $\4J_e$ and $\4J_m$ represent electric and magnetic sources, respectively. Although Maxwell's equations require $\4J_m=0$, \sl virtual \rm magnetic sources will be needed in the general formulation of Huygens' principle. As we shall see, this will not violate the prohibition of magnetic sources in nature.

To solve \eq{Max1} for $\3F$, apply $\4D$:
\begin{align}\lab{wave2}
\Box\3F=\4D\bbar D\3F=\4D\4J\=\4G
\end{align}
where
\begin{align}\lab{dJ}
\4G=(\pl_t-\grad)(\r-\3J)=(\pl_t\r+\div\3J)+(i\curl\3J-\grad\r-\pl_t\3J).
\end{align}
Since the left side of \eq{wave2} is a pure vector field, the scalar component of the right side of \eq{dJ} must vanish. This gives the continuity equation
\begin{gather}\lab{cont}
\la\4G\ra_s=\pl_t\r+\div\3J=0\\
\r=\r_e+i\r_m,\  \3J=\3J_e+i\3J_m,\nt
\end{gather}
whose real and imaginary parts state that electric and magnetic charge are conserved.
Assuming the initial condition $\3F(\3x,-\8)=\30$, we obtain the unique causal solution
\begin{align}\lab{sol2}
\3F=P*\4G=P*(\4D\4J)=\4D\,(P*\4J),
\end{align}
where the last equality follows because $\Box$ commutes with $\bbar D$ and $\4D$, and
\begin{align}\lab{dJ1}
\4G=-\grad\r-\pl_t\3J+i\curl\3J
\end{align}
by \eq{dJ} and \eq{cont}. Thus we can obtain $\3F$ in two ways: by propagating the source $\4G$, or by using the right side of \eq{sol2}:
\begin{align}\lab{FDA}
\3F\0x=\4D\4A\0x
\end{align}
where the Pauli field
\begin{align}\lab{APJ}
\4A=P*\4J=\F-\3A,\ \ \hbox{with}\ \  \F=P*\r\ \ \hbox{and}\ \ \3A=P*\3J,
\end{align}
representing the 4-potential, is the causal solution of the wave equation
\begin{align}\lab{DAJ}
\Box\4A=\bbar D\4D\4A=\bbar D\3F=\4J.
\end{align}
In fact, since $\3F$ is a pure vector field,
\begin{align}\lab{sol3}
\3F=\4D\4A=(\pl_t \F+\div\3A)-\grad \F-\pl_t\3A+i\curl\3A 
\end{align}
shows that $\4A$ satisfies the \sl Lorenz gauge condition\rm\,\footnote{Due to L V Lorenz and not H A Lorentz; see \ci{B99}. Evidently the Lorenz gauge is selected by causality, although non-causal gauges like the Coulomb gauge are admissible since the potentials are themselves unobservable in classical electrodynamics.
}
\begin{align}\lab{lor}
\pl_t \F+\div\3A=0.
\end{align}
If $\4J$ is complex as in \eq{Jem}, then so is $\4A$:
\begin{align}\lab{Aem}
\4A=\4A_e+i\4A_m&&\4A_e=\F_e-\3A_e&&\4A_m=\F_m-\3A_m.
\end{align}
The Maxwell field with electric and magnetic sources is then given by
\begin{align}\lab{EHA}
\3E&=\re\3F=-\grad \F_e-\pl_t\3A_e-\curl\3A_m\\
\3B&=\im\3F=-\grad \F_m-\pl_t\3A_m+\curl\3A_e.\nt
\end{align}
Of course, the homogeneous Maxwell equations require $\4J_m\!=\4A_m\!=0$. But the expressions  \eq{EHA} with a \sl virtual \rm magnetic 4-potential $\4A_m$ will be used to formulate Huygens' principle.

\section{Huygens' principle in electrodynamics}\lab{S:Huygens}

The assumption that $\4J\0x$ is compactly supported was made for convenience and can be relaxed. While it is reasonable to assume that the \sl spatial \rm support of $\4J$ is bounded at any time, we want to allow sources persisting in time, for example a set of charged particles following world lines or extended charged systems evolving in time. This includes, among other things, time-harmonic systems. The above results remain valid provided the integrals converge.

Let the sources be \sl spatially \rm bounded. To simplify the analysis, assume that the spatial support of $\4J\xt$ is contained in the interior of a closed surface $S\subset\rr3$ at all times $t$.\footnote{This will be generalized to sources in arbitrary motion in Section \ref{S:moving} by allowing $\l$ to depend on time. Here we assume a fixed surface $S$, as is commony done in the derivation of Huygens' principle.
} 
We assume that $S$ is a smooth manifold, at least of class $C^2$. Denote the exterior of $S$ by $E$ and its interior by $I$. Both $E$ and $I$ are taken to be \sl open \rm sets, so that $\rr3$ is the disjoint  union
\begin{align*}
\rr3=E\cup S\cup I.
\end{align*}
Let $\l\ox$ be a $C^2$ function such that\footnote{An example of a function with these properties is
\begin{align*}
\l\ox=\begin{cases}
\ \ d\ox,&\3x\in E\\ \qq\qq 0,&\3x\in S\\-d\ox,&x\in I,
\end{cases}
\end{align*}
where $d\ox$ is the shortest distance from $\3x$ to $S$.
}
\begin{align}\lab{EE}
\3x\in E&\imp\l\ox>0\\
\3x\in S&\imp \l=0\ \ \hbox{and}\ \  |\grad\l|=1\nt\\
\3x\in I&\imp\l\ox<0.\nt
\end{align}
The \sl characteristic functions \rm $\c\6E$ and $\c\6I$ of $E$ and $I$ may be written in terms of the Heaviside step function $H$ as\footnote{For $\3x\in S$, we define $\c\6E\ox=\c\6I\ox=1/2$; but this singular case will not be needed since it does not affect $\c\6E$ and $\c\6I$ as \sl distributions. \rm
}
\begin{align*}
\c\6E\ox&=H(\l\ox)=\begin{cases}
1,&\!\!\!\! \3x\in E\\0,&\!\!\!\! \3x\in I
\end{cases} \qq\qq\qq
\c\6I\ox=H(-\l\ox)=\begin{cases}
0,&\!\!\!\! \3x\in E\\ 1,&\!\!\!\! \3x\in I.
\end{cases}
\end{align*}
Define the distributional vector field
\begin{align}\lab{N}
\3N\ox\=\grad\c\6E\ox=-\grad\c\6I\ox=\d\6S\ox\3n\ox
\end{align}
where
\begin{align*}
\3n\ox=\grad\l\ox\ \ \hbox{and}\ \ \d\6S\ox=H'(\l\ox)=\d(\l\ox).
\end{align*}
Thus $\3n$ is the \sl outward \rm unit normal on $S$ and $\dd^3\3x\,\d\6S\ox$ is the 2D area measure on $S$, regarded as a singular 3D measure:
\begin{align}\lab{dSt}
\dd^3\3x\,\d\6S\ox=\dd S\ox.
\end{align}
\bf Remark. \rm Since the characteristic function $\c\6E\ox$ does not depend on the choice of $\l$, neither does the distributional field $\3N=\grad\c\6E$. The introduction of $\l$ is merely a \sl convenience \rm which helps clarify the concepts by using the relation $H'=\d$. Similar remarks will apply when $\l\xt$ is time-dependent, allowing for moving boundaries.

Let $\3F'$ be an \sl interior \rm field whose source is supported in $E$ at all times, \ie 
\begin{align}\lab{DF'}
\4J'\=\bbar D\3F',\qq \supp_{\3x}\4J'\xt\subset E\ \forall t,
\end{align}
where $\supp_\3x$ denotes \sl spatial \rm support. Since $E$ is open and the support of $\4J'$ is by definition closed, it must actually be contained in some closed set $V\subset E$. Hence $\3F'$ is defined and sourceless in an open neighborhood of $S$ as well as in the interior region $I$. Thus both $\3F$ and $\3F'$ are defined and sourceless on a neighborhood of $S$.

We shall construct a field $\3F\9S$ whose sources are concentrated on $S$ at all times and which coincides with the given field $\3F$ in $E$ and with $\3F'$ in $I$. These two partial fields are `glued' into a single field defined by
\begin{align}\lab{partn}
\3F\9S\0x=\c\6E\ox\3F\0x+\c\6I\ox\3F'\0x,\qq x=\xt\in\rr4,
\end{align}
and the source of $\3F\9S$ is \sl defined \rm by applying $\bbar D$ in a distributional sense:
\begin{align}\lab{JS}
\4J\9S=\r\9S-\3J\9S\=\bbar D\3F\9S.
\end{align}
Note that $\bbar D\c\6E=\grad\c\6E=\3N$ and
\begin{align*}
\bbar D(\c\6E\3F)&=(\bbar D\c\6E)\3F+\c\6E\bbar D\3F=\3N\3F+\c\6E\bbar D\3F.
\end{align*}
Similarly, since $\bbar D\c\6I=\grad\c\6I=-\grad\c\6E=-\3N$,
\begin{align*}
\bbar D(\c\6I\3F')=-\3N\3F+\c\6I\bbar D\3F'.
\end{align*}
Therefore
\begin{align}\lab{JS0}
\4J\9S=\3N\3F^j+\c\6E\4J+\c\6I\4J'
\end{align}
where
\begin{align}\lab{Fj}
\3F^j=\3F-\3F'=\3E^j+i\3B^j&&
\3E^j=\3E-\3E',\  \3B^j=\3B-\3B'
\end{align}
is the \sl jump discontinuity \rm across $S$. 
Since $\4J$ is supported in $I$ and $\4J'$ is supported in $E$, we have the global identities
\begin{align*}
\c\6E\4J\=0\ \ \hbox{and}\ \ \c\6I\4J'\=0.
\end{align*}
Hence
\begin{align}\lab{JS1}
\4J\9S=\3N\3F^j=\3N\cdot\3F^j+i\3N\times\3F^j=\d\6S(\3n\cdot\3F^j+i\3n\times\3F^j)
\end{align}
is a distributional charge-current density supported spatially on $S$, with a \sl surface charge-current density \rm $(\s,\3K)$ given by
\begin{align}\lab{rJ}
\r\9S=\d\6S\s&\qq\hbox{where}\ \ \s=\3n\cdot\3F^j\\
\3J\9S=\d\6S\3K&\qq\hbox{where}\ \ \3K=-i\3n\times\3F^j.\nt
\end{align}
Like $\4J$, $\4J\9S$ satisfies the distributional continuity equation
\begin{align}\lab{cons}
\la\4D\4J\9S\ra_s=\pl_t\r\9S+\div\3J\9S=0,
\end{align}
which states that charge, now restricted to flow on $S$, is conserved. 

Note that even though $\4J$ is real, $\4J\9S$ is in general \sl complex, \rm  consisting of electric and magnetic sources on $S$:
\begin{align}\lab{JSem}
\4J\9S=\4J\9S_e+i\4J\9S_m&&\4J\9S_e=\r\9S_e-\3J\9S_e&&\4J\9S_m=\r\9S_m-\3J\9S_m
\end{align}
with
\begin{align}
\r\9S_e&=\d\6S\s_e,\ \  \s_e=\3n\cdot\3E^j&&
\3J\9S_e=\d\6S\3K_e,\ \ \ \3K_e=\3n\times\3B^j\lab{surfe}\\
\r\9S_m&=\d\6S\s_m,\ \s_m=\3n\cdot\3B^j&&
\3J\9S_m=\d\6S\3K_m,\ \3K_m=-\3n\times\3E^j.\lab{surfm}
\end{align}
If we wish to construct a physically realizable surface source $\4J\9S$, then the
absence of magnetic monopoles requires it to be real:
\begin{align}\lab{hom1}
\4J\9S_m=0\iff\3n\cdot\3B^j=0 \ \ \hbox{and}\ \ \3n\times\3E^j=\30\ \ \hbox{on}\ \ S.
\end{align}
That is, the normal component of $\3B\9S$ and tangential components of $\3E\9S$ must be continuous across $S$. It can be shown\footnote{This follows in the frequency domain from  Equation (6.38) in \ci{CK92}. 
}
that the scalar condition follows from the vector condition and Maxwell's homogeneous vector equation. Since we are free to choose any sourceless interior field $\3F'$, \eq{surfe} and \eq{hom1} can be viewed as a set of \sl boundary conditions \rm for $(\3E',\3B')$ with $(\3E,\3B)$ given. Thus we look for an interior field $\3F'=\3E'+i\3B'$ such that
\begin{gather}
\pl_t\3F'+i\curl\3F'=\30\ \ \hbox{in}\ \ I\nt \\
\3n\times\3E'=\3n\times\3E\ \ \hbox{and}\ \  \3n\cdot\3B'=\3n\cdot\3B\ \ \hbox{on}\ \ S.\lab{BC}
\end{gather}
(Recall that $\3F$ and $\3F'$ extend as sourceless fields to a neighborhood of $S$.)
This boundary-value problem has a unique solution if $\3F$ is continuous in an open neighborhood of $S$, which will be the case if $\4J$ is continuous in time.\footnote{This is a sufficient but not necessary condition, as follows from the properties of the propagator \eq{P}.
Due to the factor $\d(t-r)$, the spread of $\4J$ in both time and space tends to smooth $\3F$.
}
(Recall that we have also assumed $S$ to be of class $C^2$.) For this unique interior field, \eq{surfe} and \eq{hom1} are the \sl jump conditions \rm on the interface between the interior and exterior regions \ci[pp 16--18]{J99}. 

If the interior field does \sl not \rm satisfy \eq{BC}, then the Huygens representations we are developing, although useful \sl mathematically \rm for expressing the given `real' field $\3F$ (\ie with $\4J_m=0)$ by a surface integral, cannot be realized \sl physically \rm by actual surface sources. This is what was meant by saying that the magnetic source $\4J\9S_m$ is \sl virtual. \rm In either case, we now derive the Huygens representations.

By \eq{APJ} and \eq{dSt}, the 4-potential $\4A\9S=P*\4J\9S$ for $\3F\9S$ is given by
\begin{align}\lab{AS}
\4A\9S\0x&=\int_\rr4\dd^4x'\,P(x-x')\4J\9S(x')=\int\dd^3\3x'\,\frac{\lb\4J\9S\rb}{4\p r}
\end{align}
where
\begin{align*}
r=|\3x-\3x'|\ \ \hbox{and}\ \ [\4J\9S](x,\3x')=\4J\9S(\3x', t-r)
\end{align*}
denotes the retarded source. Hence
\begin{align}\lab{FAS}
 \F\9S\0x&=\int\dd S(\3x')\,\frac{\bh n\cdot[\3F^j]}{4\p r}\\
\3A\9S\0x&=-i\int\dd S(\3x')\,\frac{\3n\times[\3F^j]}{4\p r}.\nt
\end{align}
Explicitly, the electric and magnetic 4-potentials are
\begin{align}\lab{FAS1}
\F\9S_e\0x&=\int\dd S(\3x')\,\frac{\bh n\cdot[\3E^j]}{4\p r}\\
\3A\9S_e\0x&=\int\dd S(\3x')\,\frac{\3n\times[\3B^j]}{4\p r}\nt\\
\F\9S_m\0x&=\int\dd S(\3x')\,\frac{\bh n\cdot[\3B^j]}{4\p r}\nt\\
\3A\9S_m\0x&=-\int\dd S(\3x')\,\frac{\3n\times[\3E^j]}{4\p r}.\nt
\end{align}
Although the integrations are formally over $\rr3$, they reduces to surface integrals over $S$ by \eq{dSt}. We therefore have the following result.

{\thm The field $\3F\9S=\3E\9S+i\3B\9S$ radiated by the surface charge-current density $\4J\9S=\r\9S-\3J\9S$ on $S$ is given by surface integral
\begin{align}\lab{SC0}
\3F\9S=-\grad\F\9S-\pl_t\3A\9S+i\curl\3A\9S
\end{align}
or
\begin{align}\lab{SC}
\3E\9S=-\grad\F\9S_e-\pl_t\3A\9S_e-\curl\3A\9S_m\\
\3B\9S=-\grad\F\9S_m-\pl_t\3A\9S_m+\curl\3A\9S_e\nt
\end{align}
where the surface potentials are given by {\rm\eq{FAS1}} in terms of the retarded jump discontinuities $(\3E^j, \3B^j)$ between the exterior and interior fields across $S$. This representation is valid both in the exterior region $E$, where $\3F\9S=\3F$, and in the interior region $I$, where $\3F\9S=\3F'$.
}\lab{T:Huy}

The \sl Stratton-Chu equations \rm \ci[page 32]{HY99} are a special case of \eq{SC} obtained by choosing $\3F'=\30$ and assuming that $\3x\in E$. Since $\3F'=\30$ does generally not satisfy the boundary conditions \eq{BC}, the Stratton-Chu formulation of Huygens' principle requires virtual magnetic sources on $S$. However, if $\3F'$ is chosen to be the unique solution of \eq{BC}, the Stratton-Chu equations reduce to the simpler expressions
\begin{align}\lab{SC1}
\3E\9S&=-\grad\F\9S_e-\pl_t\3A\9S_e && \3B\9S=\curl\3A\9S_e.
\end{align}

Returning to the general case \eq{FAS} and \eq{SC0}, note that $\3A\9S$ involves only the tangential components of $\3F^j$ on $S$ while $\F\9S$ involves only the normal components. The latter can be eliminated as follows. Begin with 
\begin{align*}
\pl_t\3F\9S&=-i\curl\3F\9S-\3J\9S
=-i\curl(i\curl\3A\9S-\grad\F\9S-\pl_t\3A\9S)-\3J\9S\\
&=\curl\curl\3A\9S+i\curl\pl_t\3A\9S+i\3N\times\3F^j.
\end{align*}
This involves only the tangential component $\3n\times\3F^j$ of $\3F^j$ on $S$, and  it can be integrated using the initial condition $\3A\9S(\3x,-\8)=\30$ to obtain
\begin{gather}\lab{KF0}
\3F\9S=\curl\curl\pl_t\inv\3A\9S+i\curl\3A\9S+i\3N\times\pl_t\inv\3F^j\\
\ \ \hbox{where}\ \ \pl_t\inv\3A\9S\xt=\int_{-\8}^t\dd t'\,\3A\9S(\3x, t').\nt
\end{gather}
This is a generalization of \sl Kottler-Franz equations \rm \ci[page 34]{HY99}, obtained by choosing $\3F'=\30$ and assuming that $\3x\in E$:
\begin{align}\lab{KF}
\3E\9S=\curl\curl\pl_t\inv\3A\9S_e-\curl\3A\9S_m\\
\3B\9S=\curl\curl\pl_t\inv\3A\9S_m+\curl\3A\9S_e.\nt
\end{align}
Like the Stratton-Chu equations, \eq{KF0} and \eq{KF} involve virtual magnetic sources on $S$. If we assume that the interior field satisfies the physical boundary conditions \eq{BC}, then \eq{KF} simplify to
\begin{align}\lab{KF1}
\3E\9S=\curl\curl\pl_t\inv\3A\9S_e&& \3B\9S=\curl\3A\9S_e.
\end{align}

\bf Remark. \rm Equations \eq{KF0} and \eq{SC0}, unlike the Stratton-Chu and Kottler-Franz equations, are \sl global. \rm They remain valid when $\3x\in I$ (where $\3F\9S=\3F'$) and, in a distributional sense, even when $\3x\in S$, as indicated by the last term in \eq{KF0} which is missing in \eq{KF}. Consequently, they also solve the \sl interior problem, \rm where we are given a source $\4J'$ with spatial support in $E$ and required to find its field in $I$ in terms of an equivalent source on $S$. This is useful for describing \sl reception. \rm

\section{Moving sources}\lab{S:moving}

The above can be generalized to sources in arbitrary motion simply by letting $\l$ depend on time, so that the 2D surface
\begin{align*}
S_t=\{\3x: \l\xt=0\}\subset\rr3
\end{align*}
enclosing the source $\4J\xt$ at time $t$ is time-dependent. The 3D hypersurface
\begin{align*}
\2S=\{x=\xt:\l\0x=0\}\subset\rr4
\end{align*}
now represents the \sl history \rm of $S_t$. It is the oriented boundary separating the exterior and interior spacetime regions:
\begin{align*}
\2E=\{x:\l\0x>0\},\qq \2I=\{x:\l\0x<0\},\qq \2S=\pl\2I=-\pl\2E.
\end{align*}
Define the Huygens field
\begin{align}\lab{partn2}
\3F\9{\2S}\0x=\c\6{\2E}\0x\3F\0x+\c\6{\2I}\0x\3F'\0x
\end{align}
where
\begin{align*}
\c\6{\2E}\0x=H(\l\0x)\ \ \hbox{and}\ \ \c\6{\2I}\0x=H(-\l\0x)=1-\c\6{\2E}\0x
\end{align*}
are the characteristic functions of $\2E$ and $\2I$ in spacetime.
Then the same arguments as above give
\begin{align}\lab{JS2}
\4J\9{\2S}\=\bbar D\3F\9{\2S}=(\bbar D\c\6{\2E})\3F^j=\d\6{\2S}(\dot\l\3F^j+\3n\cdot\3F^j+i\3n\times\3F^j)
\end{align}
with charge- and current distributions
\begin{align}\lab{rJS1}
\r\9{\2S}=\d\6{\2S}\,\3n\cdot\3F^j,\qq \3J\9{\2S}=-\d\6{\2S}(\dot\l\3F^j+i\3n\times\3F^j),
\end{align}
where $\dot\l=\pl_t\l$ and
\begin{align}\lab{dS2}
\d\6{\2S}\0x=\d(\l\0x)
\end{align}
is the distribution supported on $\2S$ representing the measure
\begin{align*}
\dd^4 x\,\d\6{\2S}\0x=\dd t\,\dd^3\3x\,\d(\l\xt)=\dd t\,\dd S_t\ox.
\end{align*}
The term $-\d\6{\2S}\dot\l\3F^j$ in \eq{rJS1} is a \sl drag current \rm generated by the motion of $S_t$. Since
\begin{align*}
-\dot\l\3F^j=\dot\l\3n\times(\3n\times\3F^j)-\dot\l\3n(\3n\cdot\3F^j),
\end{align*}
it has tangential and normal components. Equations \eq{FAS} generalize to
\begin{align}\lab{FAS2}
\F\9{\2S}\0x&=\int_\rr4\dd^4x'\,\d\6{\2S}(x')P(x-x')\3n(x')\cdot\3F^j(x')\\
\3A\9{\2S}\0x&=-\int_\rr4\dd^4x'\,\d\6{\2S}(x')P(x-x')\LB\dot\l(x')\3F^j(x')+i\3n(x')\times\3F^j(x')\RB\nt
\end{align}
The electric and magnetic 4-potentials are the real and imaginary parts, and the Stratton-Chu equations for a moving surface are obtained exactly as in \eq{SC}. 

Again, the source $\4J\9{\2S}$ is virtual in general. To make it real, hence realizable as a physical surface charge-current density, we must enforce the boundary conditions
\begin{align}\lab{BCt}
\3n\cdot\3B^j=0\ \ \hbox{and}\ \ \dot\l\3B^j+\3n\times\3E^j=0.
\end{align}
Note that the vector condition implies the scalar condition if $\dot\l\ne 0$. For $\dot\l=0$, this can be proved by letting $S_t$ acquire a small velocity $\dot\l$ at $\3x$ and then taking the limit
$\dot\l\to0$.

\section{Partitions of unity and energy flow}\lab{S:part}

The expression \eq{partn2} defines a global field $\3F\9{\2S}$ using a \sl partition \rm of spacetime into the exterior and interior regions separated by the interface $\2S$:
\begin{align*}
\rr4=\2E\cup\2S\cup\2I.
\end{align*}
When the sources (transmitters, scatterers, and receivers) form several \sl clusters, \rm it is useful to surround each cluster by a closed surface. Let us therefore start with a finite (or even infinite) partition of spacetime into open cells $\2E_k$ with characteristic functions
\begin{align*}
 \c_k\0x=\begin{cases}
1,&x\in E_k\\\frac12,&x\in\pl\2E_k\\
0,& x\in E_l\  \ l\ne k
\end{cases}
\end{align*}
which implies\footnote{In mathematics, a set of functions with the property \eq{PU} is called a \sl partition of unity, \rm although the functions are usually assumed to be differentiable. That the characteristic functions are discontinuous is not a problem, as we have seen, provided we view them as distributions when applying derivatives. The values of $\c_k$ on $\pl\2E_k$ don't actually matter since \eq{PU} still holds \sl almost everywhere. \rm
}
\begin{align}\lab{PU}
\sum_k\c_k\0x\=1.
\end{align}
This can be used to define a global field $\3F$ from local fields $\3F_k$ by
\begin{align}\lab{partn3}
\3F\0x=\sum_k\c_k\0x\3F_k\0x,
\end{align}
where the superscript $\2S$ has been dropped. We assume that $\3F_k$ is sourceless in an open spacetime region $\5O_k$ containing the \sl closure \rm of $\2E_k$, thus extending beyond its boundary. (The values of $\3F_k$ outside of $\5O_k$ don't matter due to the factor $\c_k$ in \eq{partn3}.) Since
\begin{align*}
\c_k\bbar D\3F_k=0\qq\forall k,
\end{align*}
the source of $\3F$ is the distribution
\begin{align}\lab{JJ0}
\4J\=\bbar D\3F
=\sum_k(\bbar D\c_k)\3F_k=\sum_k\LB\grad\c_k\cdot\3F_k+\dot\c_k\3F_k+i\grad\c_k\times\3F_k\RB,
\end{align}
giving the surface charge-current density
\begin{align}\lab{JJ}
\r=\sum_k\grad\c_k\cdot\3F_k\ \ \hbox{and}\ \ 
\3J=-\sum_k\LB\dot\c_k\3F_k+i\grad\c_k\times\3F_k\RB.
\end{align}
Since $\bbar D\c_k$ is supported on $\pl\2E_k$, $\4J$ is supported on the boundary\footnote{Since $\c_k$ decreases from 1 to 0 as we leave $\2E_k$,  $\pl\2E_k$ is oriented by the unit normal pointing into its \sl interior. \rm Therefore each interface $\pl\2E_k\cap\pl\2E_l$ between adjoining regions occurs \sl twice \rm in \eq{cell}, with opposite orientations. Hence the oriented sum  (\sl chain\rm) $\sum_k\pl\2E_k$ vanishes but the set-theoretic union $\2S$ does not.
}
\begin{align}\lab{cell}
\2S=\bigcup_k\pl\2E_k.
\end{align}
Furthermore, since
\begin{align}\lab{Ekl}
x\in\2E_k\cup\2E_l\cup(\pl\2E_k\cap\pl\2E_l) \imp \bbar D(\c_k+\c_l)=0,
\end{align}
$\4J$ depends only on the jump fields
\begin{align*}
\3F^j_{kl}=\3F_k-\3F_l,\qq x\in\2E_k\cap\2E_l
\end{align*}
across the interfaces between adjoining regions $\2E_k,\2E_l$. Thus
\begin{align}\lab{Huy3}
\3F=\4D\4A\ \ \hbox{where}\ \ \4A=P*\4J
\end{align}
gives a generalized Huygens representation of $\3F$ in terms of sources supported on $\2S$.
Furthermore, the \sl projection property \rm\footnote{Equation \eq{proj} fails \sl numerically \rm on $\pl\2E_k\cap\pl\2E_l$ where $\c_k\0x=\c_l\0x=1/2$, but it holds \sl weakly, \rm in the sense of distributions, \ie 
\begin{align*}
\int\dd^4 x\,\c_l\0x\c_m\0x f\0x=\d_{lm}\int\dd^4x\,\c_l\0x f\0x
\end{align*}
for any continuous function $f$ with compact support or rapid decay (needed when $\2E_k$ or $\2E_l$ are unbounded).
}
\begin{align}\lab{proj}
\c_l\0x\c_m\0x=\d_{lm}\c_l\0x
\end{align}
implies that quadratic expressions in $\3F$ have similar partitions. For example, the scalar Lorentz invariant\footnote{In \ci{K11a} it was shown that $|\3F^2|$ is related to the \sl electromagnetic inertia \rm and the \sl reactive energy \rm of the field.
}
\begin{align*}
\3F^2=\3F\cdot\3F=\3E^2-\3B^2+2i\3E\cdot\3B
\end{align*}
has the local partition
\begin{align*}
\3F^2=\sum_k\c_k\3F_k^2,
\end{align*}
and  the EM  energy-momentum density
\begin{gather}\lab{ES}
\4S\=\frac12\3F\3F^*=\frac12\LB\3F\cdot\3F^*+i\3F\times\3F^*\RB=U+\3S\\
U=\frac12\lp\3E^2+\3B^2\rp,\qq  \3S=\3E\times\3B,\nt
\end{gather}
(where $\3F^*$ is the ordinary complex conjugate of $\3F\in\cc3$)
has the local partition
\begin{align*}
\4S=\sum_k\c_k\4S_k\qqq U=\sum_k\c_kU_k\qqq\3S=\sum_k\c_k\3S_k.
\end{align*}
Hence the local \sl power density \rm (rate of increase of energy density) is
\begin{align*}
\la\bbar D\4S\ra_s&=\dot U+\div\3S=\sum_l\LB\dot\c_k U_k+\grad\c_k\cdot\3S_k\RB
+\sum_k\c_k\LB\dot U_k+\div\3S_k\RB.
\end{align*}
Since $\3F_k$ is sourceless in $\5O_k$, it follows from Poynting's theorem that
\begin{align*}
\dot U_k+\div\3S_k=0\ \ \hbox{in}\ \ \5O_k
\end{align*}
and thus
\begin{align}\lab{Poyn1}
\dot U+\div\3S=\sum_k\LB \dot\c_kU_k+ \grad\c_k\cdot\3S_k\RB.
\end{align}
Here $\dot\c_kU_k$ is the rate of increase in the energy density coming into $\2E_k$ due the motion of the boundary $\pl\2E_k$, and $ \grad\c_k\cdot\3S_k$ is that due to the incoming momentum flowing through $\pl\2E_k$. By \eq{Ekl}, the right side of \eq{Poyn1} involves only the \sl differences \rm
\begin{align*}
U_{kl}^j=U_k-U_l\ \ \hbox{and}\ \  \3S_{kl}^j=\3S_k-\3S_l\ \ \hbox{on}\ \ \pl\2E_k\cap\pl\2E_l.
\end{align*}
Since the general partition \eq{partn3} allows arbitrary choices of sourceless fields $\3F_k$ in domains $\5O_k$ containing the closure of $\2E_k$, these differences need not vanish. Physically, this means that energy must be pumped in or out of the boundary $\2S$ to maintain these fields.
By enforcing boundary conditions on any interface $\pl\2E_k\cap\pl\2E_l$, the corresponding terms can be made to vanish. But then that interface can be removed, thus merging the two cells into one. 

It is instructive to confirm \eq{Poyn1} using the expression \eq{JJ} for the surface current $\4J$.
Recall that $\4J$ is generally complex, including magnetic as well as electric sources:
\begin{align*}
\4J=\4J_e+i\4J_m.
\end{align*}
The generalized Poynting theorem for a complex surface current density is derived by applying the distributional Maxwell equations
\begin{align*}
\pl_t\3F+i\curl\3F=-\3J&& \pl_t\3F^*-i\curl\3F^*=-\3J^*
\end{align*}
to
\begin{align*}
\pl_t U+\div\3S
=\frac12\LB\pl_t\3F\cdot\3F^*+\3F\cdot\pl_t\3F^*+i\curl\3F\cdot\3F^*-i\3F\cdot\curl\3F^*\RB,
\end{align*}
which gives
\begin{align}\lab{Poyn}
\pl_t U+\div\3S=-\frac12\lp\3J\cdot\3F^*+\3J^*\cdot\3F\rp=-\3J_e\cdot\3E-\3J_m\cdot\3B.
\end{align}
The right side is, like $\3J$, a distribution supported on $\2S$. The partitions
\begin{align*}
 -\3J=\sum_k\LB\dot\c_k\3F_k+ i\grad\c_k\times\3F_k\RB && \3F=\sum_l\c_l\3F_l
\end{align*}
give
\begin{align*}
-\3J\cdot\3F^*
&=\sum_{kl}\LB \c_l\dot\c_k\3F_k\cdot\3F_l^*+ i\c_l\grad\c_k\times\3F_k\cdot\3F_l^*\RB.\end{align*}
Using $\grad\c_k\times\3F_k\cdot\3F_l^*=\grad\c_k\cdot\3F_k\times\3F_l^*$, \eq{Poyn} gives
\begin{align*}
\pl_t U+\div\3S
&=\frac12\sum_{kl}\LB (\c_l\dot\c_k+\c_k\dot\c_l)\3F_k\cdot\3F_l^*+ i(\c_l\grad\c_k+\c_k\grad\c_l)\cdot\3F_k\times\3F_l^*\RB.
\end{align*}
But the projection property \eq{proj} implies the distributional identities
\begin{align*}
\c_l\dot\c_k+\c_k\dot\c_l=\d_{kl}\dot\c_k\ \ \hbox{and}\ \  \c_l\grad\c_k+\c_k\grad\c_l=\d_{kl}\grad\c_k,
\end{align*}
therefore
\begin{align*}
\pl_t U+\div\3S=\sum_k\LB \dot\c_kU_k+ \grad\c_k\cdot\3S_k\RB
\end{align*}
in agreement with \eq{Poyn1}. This confirms the consistency of our computations involving bilinear distributional expressions.\footnote{Quadratic expressions in singular distributions such as $\d\0x$ do not make sense but products such as $\c_l\dot\c_k$ and $\c_l\grad\c_k$ do, due to the mild nature of the singularity of $\c_l$ (\ie its finite jump discontinuity).
}

\section{Conclusion}

We have derived a concise generalization of Huygens' principle for EM fields by combining the Pauli algebra with distribution theory. Given a closed surface $S$, we computed the surface source $\4J\9S$ on $S$ required to radiate arbitrarily given exterior and interior fields $\3F$ and $\3F'$. Then \eq{AS} gives the 4-vector potentials of $\3F$ and $\3F'$ as surface integrals over $S$. The expressions of the fields in terms of these potentials generalize the Stratton-Chu and Kottler-Franz equations. This idea was extended to a time-dependent surface, required for sources in general motion (Section \ref{S:moving}), and to multiple surfaces (Section \ref{S:part}). The latter can be applied, for example, when any number of sources, including transmitters, scatterers, and receivers, form multiple clusters in spacetime.

\section*{Acknowledgements}
I thank  David Colton  and Thorkild Hansen for helpful discussions, and Arje Nachman for his sustained support of this work, most recently through AFOSR Grant \#FA9550-08-1-0144.

\end{document}